\documentstyle[preprint,prd,aps]{revtex}

\begin{document}
\draft
\begin{titlepage}
\preprint{\vbox{\hbox{NDHU-TH-97-01} \hbox{UDHEP-11-97} 
\hbox{November 1997} }}
\title{ \large \bf Chiral Symmetry Breaking 
 in a  Uniform \\ External Magnetic Field II. \\
Symmetry Restoration at High Temperatures \\
and Chemical Potentials} 
\author{\bf D.-S. Lee$^{(a)}$, C. N. Leung$^{(b)}$
and Y. J. Ng$^{(c)}$}
\address{(a) Department of Physics, National Dong Hwa University,\\
Shoufeng, Hualien 974, Taiwan\\}
\address{(b) Department of Physics and Astronomy, 
University of Delaware, \\
Newark, DE 19716 \\}
\address{(c) Institute of Field Physics, Department of Physics 
and Astronomy,\\
University of North Carolina, Chapel Hill, NC  27599\\}

\maketitle
\begin{abstract}

Chiral symmetry is dynamically broken in quenched, ladder QED at 
weak gauge couplings when an external magnetic field is present.  
In this paper, we show that chiral symmetry is restored above a 
critical chemical potential and the corresponding phase 
transition is of first order. In contrast, the chiral symmetry 
restoration at high temperatures (and at zero chemical potential) 
is a second order phase transition.

\end{abstract}
\end{titlepage}

Do external fields affect the symmetry properties of the vacuum?
\cite{Ng}  The answer is yes; in particular, it was found that, 
in the quenched, ladder approximation of QED, chiral symmetry 
is dynamically broken at weak gauge couplings when a uniform 
magnetic field is present\cite{GMS,LNA}.  Subsequently, in a more 
detailed exposition\cite{LLN}, we show that chiral symmetry is 
restored above a critical temperature which is of the order of 
$m_0$, the dynamical fermion mass generated at zero temperature 
and zero chemical potential (see also \cite{GS}).  In this paper, 
we consider the effects of a chemical potential on the chiral 
symmetry breaking in QED induced by a  magnetic field.  We find 
that chiral symmetry is restored above a critical chemical 
potential $\mu_c$ which is proportional to $m_0$ with the 
proportional factor a function of the gauge coupling constant.  
We also find that the dynamical fermion mass $m_{\mu}$ generated 
at nonzero chemical potential increases from $m_0$ as the 
chemical potential $\mu$ is increased from zero toward $\mu_c$ 
and exhibits a discontinuous jump to $m_\mu =0$ when the critical 
chemical potential is crossed.  The order parameter for the 
phase transition, the chiral condensate $\langle \bar{\psi} 
\psi \rangle$, has the similar discontinuous behavior at 
$\mu_c$.  This indicates that the corresponding phase 
transition is of first order.  For comparison, we study 
the nature of the chiral phase transition at high 
temperature but with zero chemical potential. This phase 
transition is of second order since the dynamical fermion 
thermal mass $m_T$ as well as the corresponding chiral 
condensate approach zero continuously as the temperature 
approaches the critical temperature from below.

Unless otherwise noted, all notations are the same as those in 
Ref.\cite{LLN}, hereafter referred to as I.

The effects of a chemical potential $\mu$ can be incorporated 
into the study of chiral symmetry breaking in an external 
magnetic field at nonzero temperature\cite{LLN} by making the 
substitution\cite{Ka}:
\begin{equation}
 p_0 \rightarrow i \pi T(2l+1) - \mu, ~~~~l=0, \pm 1, \pm 2,... 
\end{equation}
for the fermion energy.  The thermal photon propagator remains 
unchanged. One can easily show that the presence of a chemical 
potential does not modify the orthogonality and completeness 
conditions\cite{OC} of the basis eigenfunctions used in our 
formalism.  It is then straightforward to follow the procedures 
described in I to obtain the gap equation for nonzero temperature 
and nonzero chemical potential:
\begin{equation}
 1 \simeq \frac{2 \alpha}{\pi} T |eH| \int_{-\infty}^{\infty} 
 dq_3 \int_0^\infty d\hat{q}_\perp^2 ~{\rm e}^{- \hat{q}_\perp^2} 
 \sum_{l} \frac{1}{Q_2^2 + 4 \pi^2 T^2 l^2}~\frac{1}{Q_1^2 + 
 [ \pi T (2 l -1) - i \mu]^2} 
\label{tmugap}
\end{equation}
where  $~Q_1^2 \equiv q_3^2 + m^2 (T,\mu)$, $~Q_2^2 \equiv q_3^2 
+ 2 |eH| \hat{q}_\perp^2$, and $m(T,\mu)$ is the infrared dynamical 
fermion mass as a function of both the temperature and the chemical 
potential.

We sum over $l$ in Eq.(\ref{tmugap}) using the Poisson sum 
formula.  The gap equation now reads
\begin{eqnarray}
 1 &~\simeq~& \frac{\alpha}{2 \pi} |eH| \int_{-\infty}
 ^\infty dq_3 \int_0^\infty d\hat{q}_\perp^2~  
 \frac{{\rm e}^{- \hat{q}_\perp^2}}{Q_1 Q_2}
\nonumber \\
 & & \cdot \left\{ Q_1 \coth(\frac{Q_2}{2T}) \left[ \frac{1}
 {Q_1^2 - (Q_2+\mu-i\pi T)^2} + \frac{1}{Q_1^2 - (Q_2 - \mu + 
 i \pi T)^2} \right] \right. 
\nonumber \\
 & & ~~~~+ \left. Q_2 \left[ \frac{\tanh( \frac{Q_1+\mu}{2T})}
 {Q_2^2-(Q_1 + \mu - i \pi T)^2} + \frac{\tanh(\frac{Q_1-\mu}{2T})}
 {Q_2^2-(Q_1 - \mu + i \pi T)^2} \right] \right\}_. 
\label{gaptu}
\end{eqnarray}
In the $\mu =0$ limit, this equation reduces to the nonzero
temperature gap equation obtained in I \cite{typo}.

Following the procedure described in Appendix B of I, one can 
obtain the chiral condensate at nonzero temperature and nonzero 
chemical potential as 
\begin{equation}
 \langle \bar{\psi} \psi \rangle_{T \mu}~\simeq~-~\frac{2 \vert 
 eH \vert}{\pi^2} T \sum_l \int_0^{\sqrt{\vert eH \vert}} dp_3 
 \frac{m(T, \mu)}{[(2l+1) \pi T + i \mu]^2 + p_3^2 + m^2(T, \mu)}_,
\label{cc}
\end{equation}
where we have cut off the momentum integral at $\sqrt{\vert eH 
\vert}$ (because the dynamical fermion mass would be vanishingly 
small at momenta above $\sqrt{\vert eH \vert}$) and ignored 
the momentum dependence in $m$.  The sum can again be evaluated 
by means of the Poisson sum formula to yield 
\begin{equation}
 \langle \bar{\psi} \psi \rangle_{T \mu}~\simeq~-~\frac{\vert eH 
 \vert}{\pi^2} m \int_0^{\sqrt{\vert eH \vert}} \frac{dp_3}
 {\sqrt{p_3^2+m^2}} S_{T \mu},
\label{cctmu}
\end{equation}
where 
\begin{equation}
 S_{T \mu}~=~\frac{1}{1+{\rm e}^{-\frac{\sqrt{p_3^2+m^2}+|\mu|}
 {T}}} - \frac{\theta(|\mu|-\sqrt{p_3^2+m^2})}{1+{\rm e}^
 {-\frac{|\mu|-\sqrt{p_3^2+m^2}}{T}}} - \theta(\sqrt{p_3^2+m^2}
 -|\mu|) \left[1 - \frac{1}{1+{\rm e}^{-\frac{\sqrt{p_3^2+m^2}
 -|\mu|}{T}}} \right]_.
\label{S}
\end{equation}
In the limit $T = \mu = 0$, $S_{00}$ equals 1 and we get ($m_0 
\equiv m(0, 0)$) 
\begin{eqnarray}
 \langle \bar{\psi} \psi \rangle_{00}&~\simeq~&-~\frac{\vert eH 
 \vert}{\pi^2} m_0 \int_0^{\sqrt{\vert eH \vert}} \frac{dp_3}
 {\sqrt{p_3^2+m_0^2}}
\nonumber \\
 &~\simeq~& -~\frac{\vert eH \vert}{2 \pi^2} m_0 \ln \left(
 \frac{\vert eH \vert}{m_0^2} \right)_,
\label{cc0}
\end{eqnarray}
in agreement with the result given by Eq.(B4) in I.

To isolate the effects of the chemical potential on chiral 
symmetry breaking in the presence of an external magnetic field, 
we now consider the $~T=0~$ limit of the above gap equation.  
We note the following simplification in this limit:
\[\begin{array}{cll}
 \coth \left( \frac{Q_2}{2T} \right) & \rightarrow & 1,  \\
 \tanh \left( \frac{Q_1+\mu}{2T} \right) & \rightarrow & 
 \theta(\mu) ~+~ \theta(-\mu)~\left[\theta(Q_1+\mu) - 
 \theta(-\mu - Q_1) \right], \\
 \tanh \left( \frac{Q_1-\mu}{2T} \right) & \rightarrow & 
 \theta(\mu)~\left[\theta(Q_1-\mu) - \theta(\mu - Q_1)\right] 
 ~+~ \theta(-\mu).
\end{array}\]
Eq.(\ref{gaptu}) is then reduced to 
\begin{eqnarray}
 1 &~\simeq~&  \frac{\alpha}{2 \pi} |eH| \int_{-\infty}^\infty 
 dq_3 \int_0^\infty d\hat{q}_\perp^2 ~
  \frac{{\rm e}^{-\hat{q}_\perp^2}}{Q_1 Q_2} \nonumber \\
 & & ~~~~\cdot \left( \frac{1}{Q_1+Q_2+ \vert \mu \vert} +
 \frac{\theta (Q_1- \vert \mu \vert)}{Q_1+Q_2-\vert \mu \vert} 
 + \frac{\theta(\vert \mu \vert -Q_1)}{Q_1-Q_2- \vert \mu \vert} 
 \right)_.
\label{gapu}
\end{eqnarray}
As a check, in the $~\mu =0~$ limit, we recover Eq.($65$) in I, 
which is the gap equation for the case of $~T=0~$ and $~\mu=0$.  
Notice that Eq.(\ref{gapu}) as well as Eq.(\ref{cctmu}) do not 
depend on the sign of $\mu$.  For ease of writing, henceforth 
we drop the magnitude notation on $\vert \mu \vert$ with the 
understanding that $\mu$ stands for the nonnegative $\vert 
\mu \vert$.

In the small $\mu$ limit where $~\mu \ll m_0$, we may 
treat the chemical potential effects as a perturbation and 
write $~m^2(0,\mu) \simeq m_0^2 + \delta m^2_{\mu}~$ with 
$~\delta m^2_{\mu} \ll m^2_0$.  This allows us to obtain
\begin{equation}
 \delta m^2_{\mu} \simeq \frac{2 I_2}{I_1} \mu^2,
\label{approxgapu}
\end{equation}
where
\begin{eqnarray}
 I_1 &~=~& \int_{-\infty}^{\infty} dq_3 \int_{0}^{\infty} 
  d\hat{q}_\perp^2~ \frac{{\rm e}^{-\hat{q}_\perp^2}}{Q^2 
  Q_2 (Q + Q_2)} \left(\frac{1}{Q} + \frac{1}{Q + Q_2} 
  \right)_, \nonumber \\
 I_2 &~=~& \int_{-\infty}^{\infty} dq_3 \int_{0}^{\infty} 
  d\hat{q}_\perp^2 ~\frac{{\rm e}^{-\hat{q}_\perp^2}} 
  {Q Q_2 (Q + Q_2)^3}_,   
\end{eqnarray}
and $~Q^2 \equiv q_3^2 + m_0^2$. Note that $I_1$ and $I_2$ 
are the same integrals as defined in Eq.($67$) of I.  They 
are both finite and positive so that $\delta m^2_{\mu}$ is 
positive.

In the large $\mu$ limit where $~\mu \gg m_0$, we may 
rewrite Eq.(\ref{gapu}) as 
\begin{eqnarray}
 1 ~&\simeq&~ - \frac{2\alpha}{\pi} |eH| \int_0^\infty 
 d\hat{q}_\perp^2 ~ {\rm e}^{-\hat{q}_\perp^2}
 \left[\int_0^{\sqrt{\mu^2-m^2}} \frac{dq_3}{Q_2}~ 
 \frac{1}{(\mu+Q_2)^2 - Q_1^2} \right.
\nonumber \\
 & & ~~~~~~~~~~- \left. \int_{\sqrt{\mu^2-m^2}}^\infty 
 \frac{dq_3}{Q_2}~\frac{1+\frac{Q_2}{Q_1}}
 {(Q_1+Q_2)^2 - \mu^2} \right]_.
\end{eqnarray}
Since both $q_3$-integrals are dominated by values of $q_3$ 
near the respective lower limit of integration, it is clear 
that the first term in the square bracket is larger than the 
second term.  Consequently the right hand side of the gap 
equation is negative and there is no solution for large $\mu$.  
This implies that there is a critical chemical potential 
above which chiral symmetry is restored.  Indeed, this 
expectation is confirmed by the result of our numerical 
study discussed below.

Being unable to solve the gap equation, Eq.(\ref{gapu}), 
analytically, we seek a numerical solution instead. It is 
convenient to introduce the following dimensionless 
variables:
\begin{equation}
 \hat{q}_3 \equiv \frac{q_3}{\sqrt{2\vert eH \vert}}_,~~~
 \hat{m}_{\mu}\equiv\frac{m(0,\mu)}{\sqrt{2\vert eH \vert}}_, 
 ~~~\hat{\mu} \equiv \frac{\mu}{\sqrt{2\vert eH \vert}}_, 
 ~~~\hat{Q}_{1,2}\equiv\frac{Q_{1,2}}{\sqrt{2\vert eH \vert}}_.
\end{equation}
In terms of these variables, Eq.(\ref{gapu}) becomes
\begin{equation}
 1 \simeq \frac{2\alpha}{\pi}~I(\hat{m}_{\mu}, \hat{\mu}),
\label{gapuless}
\end{equation}
where 
\begin{eqnarray}
 I(\hat{m}_{\mu}, \hat{\mu}) &=& \frac{1}{4} 
 \int_{0}^\infty d\hat{q}_3 \int_{0}^\infty d\hat{q}_\perp^2 
 ~~\frac{{\rm e}^{-\hat{q}_\perp^2}}{\hat{Q}_1 \hat{Q}_2} 
\nonumber \\
 & & ~~~~\cdot \left( \frac{1}{\hat{Q}_1 + \hat{Q}_2 + \hat{\mu}} 
 + \frac{\theta(\hat{Q}_1-\hat{\mu})}{\hat{Q}_1 + \hat{Q}_2 - 
 \hat{\mu}} - \frac{\theta(\hat{\mu}-\hat{Q}_1)}{\hat{\mu} - 
 \hat{Q}_1 + \hat{Q}_2} \right)_.
\label{I}
\end{eqnarray}
For various values of $~\hat{\mu}~$ and $~\hat{m}_{\mu}~$, one 
can perform the double integration numerically.  The result is 
shown in Fig. 1 where we have plotted $~I(\hat{m}_{\mu}, 
\hat{\mu})~$ as a function of $~\hat{m}_{\mu}~$ for five 
representative values of successively increasing $~\hat{\mu}~$ 
from $~\hat{\mu}=0~$ to $~\hat{\mu}=0.004243$.  The numerical 
solution for $~\hat{m}_\mu~$ for each value of $~\hat{\mu}~$ 
can be read off by equating $~I(\hat{m}_{\mu}, \hat{\mu})~$ to 
$\pi/2 \alpha$.  For illustrative purpose, we have 
used $~\alpha=\pi/30~$ in Fig. 1.  Going from Fig. 1b to Fig. 1c, 
as $\hat{\mu}$ increases, one finds two solutions for 
$\hat{m}_\mu$ instead of one.  As is clear from Fig. 1a,b,c, 
the requirement of continuity to the $~\hat{\mu}=0~$ solution 
selects the larger value of $\hat{m}_{\mu}$ as the consistent 
solution \cite{EXP}.  Increasing $~\hat{\mu}$ to $~0.004101$, 
one finds again only one solution for $~\hat{m}_{\mu}$, as 
shown in Fig. 1d.  Further increase in $\hat{\mu}$ results in 
no solution as indicated in Fig. 1e.   We  therefore conclude that 
the critical value of $~\hat{\mu}~$ is $~0.004101~$ (for $~\alpha 
=\pi/30$) above which chiral symmetry is restored.

In Fig. 2, the dimensionless dynamical fermion mass 
$~\hat{m}_{\mu}~$ is plotted as a function of the dimensionless 
chemical potential $\hat{\mu}$ for $~\alpha =\pi/30$.  We 
note that the phase transition across the critical chemical 
potential is of first order as the dynamical fermion mass 
has a discontinuity at the transition point.  The order parameter 
also exhibits such a discontinuity.  The chiral condensate at 
$~T = 0~$ and $~\mu \neq 0~$ can be obtained from Eq.(\ref{cctmu}) 
by noting that $S_{0 \mu} = \theta(\sqrt{p_3^2+m_\mu^2} - \mu)$, 
where $m_\mu$ denotes $m(0, \mu)$.  We find 
\begin{equation}
 \langle \bar{\psi} \psi \rangle_{0 \mu}~\simeq~-~\frac{\vert eH 
 \vert}{\pi^2} m_\mu \left[\theta(m_\mu-\mu) \int_0^{\sqrt{\vert 
 eH \vert}} \frac{dp_3}{\sqrt{p_3^2+m_\mu^2}} + \theta(\mu-m_\mu) 
 \int_{\sqrt{\mu^2-m_\mu^2}}^{\sqrt{\vert eH \vert}} \frac{dp_3}
 {\sqrt{p_3^2+m_\mu^2}} \right]_.
\label{ccmu}
\end{equation}
Since $~m_\mu = 0~$ above the critical chemical potential and 
$~m_\mu > \mu~$ for $~\mu < \mu_c~$ (see Fig. 2), only the first 
integral on the right hand side of Eq.(\ref{ccmu}) contributes.  
The result is 
\begin{equation}
 \langle \bar{\psi} \psi \rangle_{0 \mu}~\simeq~-~\frac{\vert eH 
 \vert}{2 \pi^2} m_\mu \ln \left(\frac{\vert eH \vert}{m_\mu^2} 
 \right)_.
\label{ccmu2}
\end{equation}
Clearly, the result in Eq.(\ref{cc0}) is recovered when $\mu = 0$.  
It is also apparent that the discontinuity in $m_\mu$ is reflected 
as a discontinuity in $\langle \bar{\psi} \psi \rangle_{0 \mu}$ at 
the critical chemical potential.

One can numerically compute the critical chemical potential 
$\hat{\mu}_c$ for different values of $\alpha$.  The result, 
as depicted by the solid curve in Fig. 3, shows that $\hat{\mu}_c$ 
increases with $\alpha$.  This is an expected behavior because 
it requires a larger chemical potential to destabilize a more 
strongly coupled chiral condensate.

A closer examination of Fig. 1 shows that, for each value of 
$~\hat{\mu}$, $~I(\hat{m}_{\mu}, \hat{\mu})~$ peaks at 
$~\hat{m}_{\mu} =\hat{\mu}$.  One can understand this feature 
qualitatively.  For a given value of $~\hat{\mu}$, let us 
consider the integral $~I(\hat{m}_{\mu}, \hat{\mu})~$ as a 
function of $~\hat{m}_{\mu}$. For $~\hat{m}_{\mu} > \hat{\mu}$, 
Eq.(\ref{I}) becomes
\begin{eqnarray}
 I(\hat{m}_{\mu}, \hat{\mu}) &=& \frac{1}{4} \int_{0}^\infty 
 d\hat{q}_3 \int_{0}^\infty d\hat{q}_\perp^2 ~~\frac
 {{\rm e}^{-\hat{q}_\perp^2}}{\sqrt{({\hat{q}_3}^2 + 
 \hat{m}_{\mu}^2)({\hat{q}_3}^2 + \hat{q}_\perp^2)}} 
\nonumber \\
 & &~~~~~\cdot \left(\frac{1}{\sqrt{{\hat{q}_3}^2 + 
 \hat{m}_{\mu}^2} + \hat{\mu} + \sqrt{\hat{q}_3^2 + 
 \hat{q}_\perp^2}} + \frac{1}{\sqrt{{\hat{q}_3}^2 + 
 \hat{m}_{\mu}^2} - \hat{\mu} + \sqrt{\hat{q}_3^2 + 
 \hat{q}_\perp^2}} \right)_.
\end{eqnarray}
Clearly, $~I(\hat{m}_{\mu}, \hat{\mu})~$ increases monotonically 
as $\hat{m}_{\mu}$ is decreased; so in the region where 
$~\hat{m}_{\mu} > \hat{\mu}$, $~I(\hat{m}_{\mu}, \hat{\mu})~$ 
reaches its maximum at $~\hat{m}_{\mu} = \hat{\mu}$.  For 
$~\hat{m}_{\mu} < \hat{\mu}$, we may examine the slope 
\begin{eqnarray}
 \frac{\partial I}{\partial \hat{m}_\mu} &=& \hat{m}_{\mu}~ 
 \int_{0}^\infty d\hat{q}_3 \int_{0}^\infty d\hat{q}_\perp^2 
 ~~\frac{{\rm e}^{-\hat{q}_\perp^2}}{\hat{Q}_2} \left\{ \frac
 {1}{\hat{Q}_2} ~\frac{1}{2 \hat{\mu} \sqrt{\hat{\mu}^2-
 \hat{m}_\mu^2}} ~\delta \left(\hat{q}_3 - \sqrt{\hat{\mu}^2-
 \hat{m}_\mu^2} \right) \right. 
\nonumber \\
 & &~~~~~ \left. -~\frac{\theta(\hat{\mu}-\hat{Q}_1)}
 {[(\hat{\mu}+\hat{Q}_2)^2 - \hat{Q}_1^2]^2} \right.
\nonumber \\ 
 & &~~~~~ \left. -~ \frac{\theta(\hat{Q}_1-\hat{\mu})}
 {[(\hat{Q}_1 + \hat{Q}_2)^2 - \hat{\mu}^2]^2} \left(1 + 
 \frac{\hat{Q}_2}{2 \hat{Q}_1} \left[\left(2+\frac{\hat{Q}_2}
 {\hat{Q}_1}\right)^2 + \left(1-\frac{\hat{\mu}^2}{\hat{Q}_1^2} 
 \right) \right] \right) \right\}_.
\label{dIdm}
\end{eqnarray}
Note that it is zero at $\hat{m}_{\mu} = 0$ and goes to infinity 
as $\hat{m}_{\mu}$ approaches $\hat{\mu}$.  Furthermore, 
$\frac{1}{\hat{m}_\mu} \frac{\partial I}{\partial \hat{m}_\mu}$
is positive at $\hat{m}_{\mu} = 0$ and the  
positive term in the integrand (the term proportional to the 
$\delta$-function) increases with increasing $\hat{m}_{\mu}$ 
while all the negative terms decrease in magnitude.  In other 
words, the slope is always positive for $\hat{m}_{\mu} < 
\hat{\mu}$.  Hence, for fixed $\hat{\mu}$, $I(\hat{m}_{\mu}, 
\hat{\mu})$ is a monotonically increasing function 
of $\hat{m}_{\mu}$ in the region $\hat{m}_{\mu} < \hat{\mu}$.  
The discontinuity of the slope at $\hat{m}_{\mu} = \hat{\mu}$ 
indicates that $~I(\hat{m}_{\mu}, \hat{\mu})~$ should peak sharply 
there, as we see from Fig. 1.  We shall exploit this fact to 
estimate analytically the critical chemical potential.

As the chemical potential reaches $\hat{\mu}_c$, 
$~I(\hat{m}_{\mu}, \hat{\mu}_c)~$ peaks at $~\hat{m}_{\mu} 
= \hat{\mu}_c~$ and $~I(\hat{\mu}_c , \hat{\mu}_c) = 
\pi/ 2\alpha~$ satisfies the gap equation, 
Eq.(\ref{gapuless}). This implies that 
\begin{equation}
 \hat{m}_{\mu}^2(\hat{\mu} = \hat{\mu}_c) = \hat{\mu}_c^2.
\end{equation}
Adopting the perturbative result in Eq.(\ref{approxgapu}), we find 
\begin{equation}
 {\mu}_c \simeq \frac{m_0}{\sqrt{1 - \frac{2 I_2}{I_1}}}_.
\label{mucrit}
\end{equation}
This estimate of $\mu_c$ is shown as the dotted curve in Fig. 3 
where the approximate analytical solution for $m_0$ (see Eq.(53) 
in I) has been used.  The good agreement with the exact numerical 
result (solid curve in Fig. 3) indicates that the analytical 
estimate of the critical chemical potential is quite robust for 
small values of $\alpha$.

As noted above, the phase transition across the critical chemical 
potential is of first order.  In I where we studied chiral symmetry 
breaking in an external magnetic field at nonzero temperature but 
with zero chemical potential, we showed that the chiral symmetry 
is restored above a critical temperature $~T_c$.  In the 
remainder of this paper, we would like to examine this phase 
transition across $T_c$ in greater detail.  With the 
introduction of the dimensionless temperature and dynamical 
fermion thermal mass:
\begin{equation}
 \hat{T} \equiv \frac{T}{\sqrt{2\vert eH \vert}}_, 
 ~~~~~ \hat{m}_{T} \equiv \frac{m(T,0)}{\sqrt{2\vert eH \vert}}_,
\end{equation}
the finite temperature gap equation can be obtained by setting 
$~\mu =0~$ in Eq.(\ref{gaptu}):
\begin{equation}
  1 \simeq \frac{2 \alpha}{\pi} ~J(\hat{m}_{T}, \hat{T}),
\label{tgap}
\end{equation}
where
\begin{eqnarray}
 J(\hat{m}_T, \hat{T}) &=&  \frac{1}{2} \int_0^\infty d\hat{q}_3 
 \int_0^\infty d\hat{q}_\perp^2~\frac{{\rm e}^{-\hat{q}_\perp^2}}
 {\sqrt{(\hat{q}_3^2+\hat{m}_T^2)(\hat{q}_3^2+\hat{q}_\perp^2)}}
\nonumber \\
 & & \!\!\!\!\!\!\!\!\!\!\!\! \cdot \left[\left(\frac{1}{1-
 {\rm e}^{-\frac{\sqrt{\hat{q}_3^2+\hat{q}_\perp^2}}{\hat{T}}}} 
 + \frac{1}{1+{\rm e}^{-\frac{\sqrt{{\hat q_3}^2+\hat{m}_T^2}}
 {\hat{T}}}} - 1 \right) \frac{\sqrt{{\hat q_3}^2+\hat{m}_T^2} 
 + \sqrt{{\hat q_3}^2+\hat{q}_\perp^2}}{(\sqrt{{\hat q_3}^2
 +\hat{m}_T^2} + \sqrt{{\hat q_3}^2+\hat{q}_\perp^2})^2 
 + \pi^2 \hat{T}^2} \right. 
\nonumber \\
 & & \!\!\!\!\!\!\!\!\!\!\!\! + \left. \left(\frac{1}{1-{\rm e}^
 {-\frac{\sqrt{{\hat q_3}^2+\hat{q}_\perp^2}}{\hat{T}}}} - 
 \frac{1}{1+{\rm e}^{-\frac{\sqrt{{\hat q_3}^2+\hat{m}_T^2}}
 {\hat{T}}}} \right)  
 \frac{\sqrt{{\hat q_3}^2+\hat{m}_{T}^{2}} - 
 \sqrt{{\hat q_3}^2+\hat{q}_\perp^2}}{(\sqrt{{\hat q_3}^2
 +\hat{m}_{T}^{2}} - \sqrt{{\hat q_3}^2+\hat{q}_\perp^2})^2 
 + \pi^2 \hat{T}^2} \right]_. 
\label{gaptless}
\end{eqnarray}
We have solved the above equation numerically. The result is 
shown in Fig. 4 where, for illustrative purpose, we have again 
used $~\alpha=\pi/30$.  The critical temperature at which the 
dynamical fermion mass vanishes is $~\hat{T}_c \approx 0.00205~$ 
for this value of $\alpha$.  As shown in Fig. 4, starting 
at $~\hat{T} = \hat{T}_c$, $\hat{m}_{T}$ increases 
as $\hat{T}$ decreases and it remains zero as $\hat{T}$ 
increases indicating that the corresponding phase transition 
is of second order.

To find out how the dynamical fermion mass varies with 
temperature near (and below) $~\hat{T}_c$, we expand 
$J(\hat{m}_{T}, \hat{T})$ around $~\hat{m}_{T}^2 = 0~$ and 
$~\hat{T} = \hat{T}_c~$ so that for small $~\hat{T}-\hat{T}_c~$ 
and small $~\hat{m}_{T}^2$, Eq.(\ref{tgap}) takes the form
\begin{equation}
 1 \simeq \frac{2 \alpha}{\pi} \left[ J(0, \hat{T}_c) + 
 \frac{\partial J (0, \hat{T}_c)}{\partial \hat{T}} 
 (\hat{T}-\hat{T}_c) + \frac{\partial J (0, \hat{T}_c)}
 {\partial \hat{m}^2_T}  \hat{m}^2_{T} \right]_.
\label{approxgap}
\end{equation}
It can be shown both analytically and numerically that 
$\frac{\partial J(0, \hat{T}_c)}{\partial \hat{T}}$ and 
$ \frac{\partial J(0, \hat{T}_c)}{\partial \hat{m}^2_T}$ 
are finite and negative.  We can understand qualitatively 
why these two partial derivatives must have the same sign.  
According to the result in Fig. 4, increasing the temperature 
reduces the dynamical fermion thermal mass.  It follows 
that raising the temperature must have the opposite 
effect on $J(\hat{m}_{T}, \hat{T})$ as lowering the 
fermion thermal mass in order that the gap equation, 
Eq.(\ref{tgap}), remains satisfied.  Since $\hat{m}^2_T =0$ 
for $\hat{T} =\hat{T}_c$ solves the gap equation, we obtain 
from Eq.(\ref{approxgap}) that 
\begin{equation}
 0 = -\frac{\partial J(0, \hat{T}_c)}{\partial \hat{T}} 
 \hat{T}_c \left(1 - \frac{\hat{T}}{\hat{T}_c} \right) 
 + \frac{\partial J(0, \hat{T}_c)}{\partial 
 \hat{m}^2_T}~\hat{m}^2_T,
\end{equation}
which indicates that $\hat{m}_T$ has the behavior 
\begin{equation}
 \hat{m}_T \sim \left(1 - \frac{T}{T_c}\right)^{\frac{1}{2}}
\label{mc}
\end{equation}
as $T$ approaches $T_c$ from below.  This translates into 
a similar behavior for the order parameter of this phase 
transition.

The chiral condensate at nonzero temperature and zero 
chemical potential can be gleaned from Eq.(\ref{cctmu}) 
to be
\begin{eqnarray}
 \langle \bar{\psi} \psi \rangle_{T0} 
 &~\simeq~& 
 -~\frac{|eH|}{\pi^2} m_T \int_0^{\sqrt{|eH|}} ~\frac{dp_3} 
 {\sqrt{p_3^2 + m_T^2}} \tanh \left(\frac{\sqrt{p_3^2+m_T^2}}
 {2T} \right) 
\nonumber \\
 &~\simeq~&
 -~\frac{\vert eH \vert}{2 \pi^2} m_T \ln \left(\frac{\vert eH 
 \vert}{m_T^2} \right)_,
\label{ccT}
\end{eqnarray}
where $m_T$ denotes $m(T, 0)$ and the last result is obtained 
by noting that the integral is dominated by large values of 
$p_3$ for which the hyperbolic tangent function in the integrand 
is approximately 1 (for $T < T_c$).  Eq.(\ref{mc}) and  
Eq.(\ref{ccT}) show that 
\begin{equation}
 \langle \bar{\psi} \psi \rangle_{T0} ~\sim~
 \vert eH \vert^{\frac{3}{2}} \left(1 - \frac{T}{T_c} \right)^
 {\frac{1}{2}} \ln \left(1 - \frac{T}{T_c} \right)^{\frac{1}{2}}
\end{equation}
as $T \rightarrow T_c^-$.

To summarize, we have considered the separate effects of 
a chemical potential and temperature on the chiral symmetry 
breaking induced by an external uniform magnetic field in 
quenched, ladder QED.  We have done so in order to highlight 
the effects of each.  Of course, it is useful and interesting 
to consider their combined effects.  Work along this direction 
is in progress. \\

\bigskip
\begin{center} 
{\bf ACKNOWLEDGEMENTS}\\
\end{center}

This work was supported in part by the U.S. Department of Energy 
under Grants No. DE-FG02-84ER40163 and DE-FG05-85ER-40219 Task A.
Part of this work was done recently when C.N.L. was visiting the 
Institute of Physics at the Academia Sinica in Taipei.  He thanks 
H.-L. Yu and other institute members there for their hospitality.  
He also thanks the National Science Council of the R.O.C. for 
support under Grant NSC86-2811-M-001-065R.  
  
\raggedbottom

\newpage
\begin{center}
\center{{\bf FIGURE CAPTIONS}}
\end{center}
\vspace*{0.7cm}

\noindent
{\bf Fig. 1a:} $~I(\hat{m}_{\mu}, \hat{\mu})~$ vs. $~\hat{m}_{\mu}~$ 
for $~\hat{\mu} =0$. $~I(\hat{m}_{\mu}, \hat{\mu})~$ peaks at 
$~\hat{m}_{\mu}=\hat{\mu} = 0$.  The solution to the gap equation 
is at $~\hat{m}_{\mu}= 0.003790~$ for $~\alpha = \pi/30$.\\

\noindent
{\bf Fig. 1b:} $~I(\hat{m}_{\mu}, \hat{\mu})~$ vs. $~\hat{m}_{\mu}~$ 
for $~\hat{\mu} = 0.001414$. $~I(\hat{m}_{\mu}, \hat{\mu})~$ peaks 
at $~\hat{m}_{\mu} = \hat{\mu} = 0.001414$.  The solution to the gap 
equation is at $~\hat{m}_{\mu} = 0.003817~$ for $~\alpha = \pi/30$.\\

\noindent
{\bf Fig. 1c:} $~I(\hat{m}_{\mu}, \hat{\mu})~$ vs. $~\hat{m}_{\mu}~$ 
for $~\hat{\mu} = 0.002828$. $~I(\hat{m}_{\mu}, \hat{\mu})~$ peaks 
at $~\hat{m}_{\mu} = \hat{\mu} = 0.002828$.  Note that there are two 
possible solutions to the gap equation at $~\hat{m}_{\mu} = 0.002715$ 
and 0.003889 for $~\alpha = \pi/30$.\\

\noindent
{\bf Fig. 1d:} $~I(\hat{m}_{\mu}, \hat{\mu})~$ vs. $~\hat{m}_{\mu}~$ 
for $~\hat{\mu} = 0.004101$. $~I(\hat{m}_{\mu}, \hat{\mu})~$ peaks 
at $~\hat{m}_{\mu} = \hat{\mu} = 0.004101$.  There is one solution to 
the gap equation at $~\hat{m}_{\mu} = \hat{\mu} = 0.004101~$ for 
$~\alpha = \pi/30$.\\

\noindent
{\bf Fig. 1e:} $~I(\hat{m}_{\mu}, \hat{\mu})~$ vs. $~\hat{m}_{\mu}~$ 
for $~\hat{\mu} = 0.004243$. $~I(\hat{m}_{\mu}, \hat{\mu})~$ peaks 
at $~\hat{m}_{\mu} = \hat{\mu} = 0.004243$.  There is no solution to 
the gap equation for $~\alpha =\pi/30$.\\

\noindent
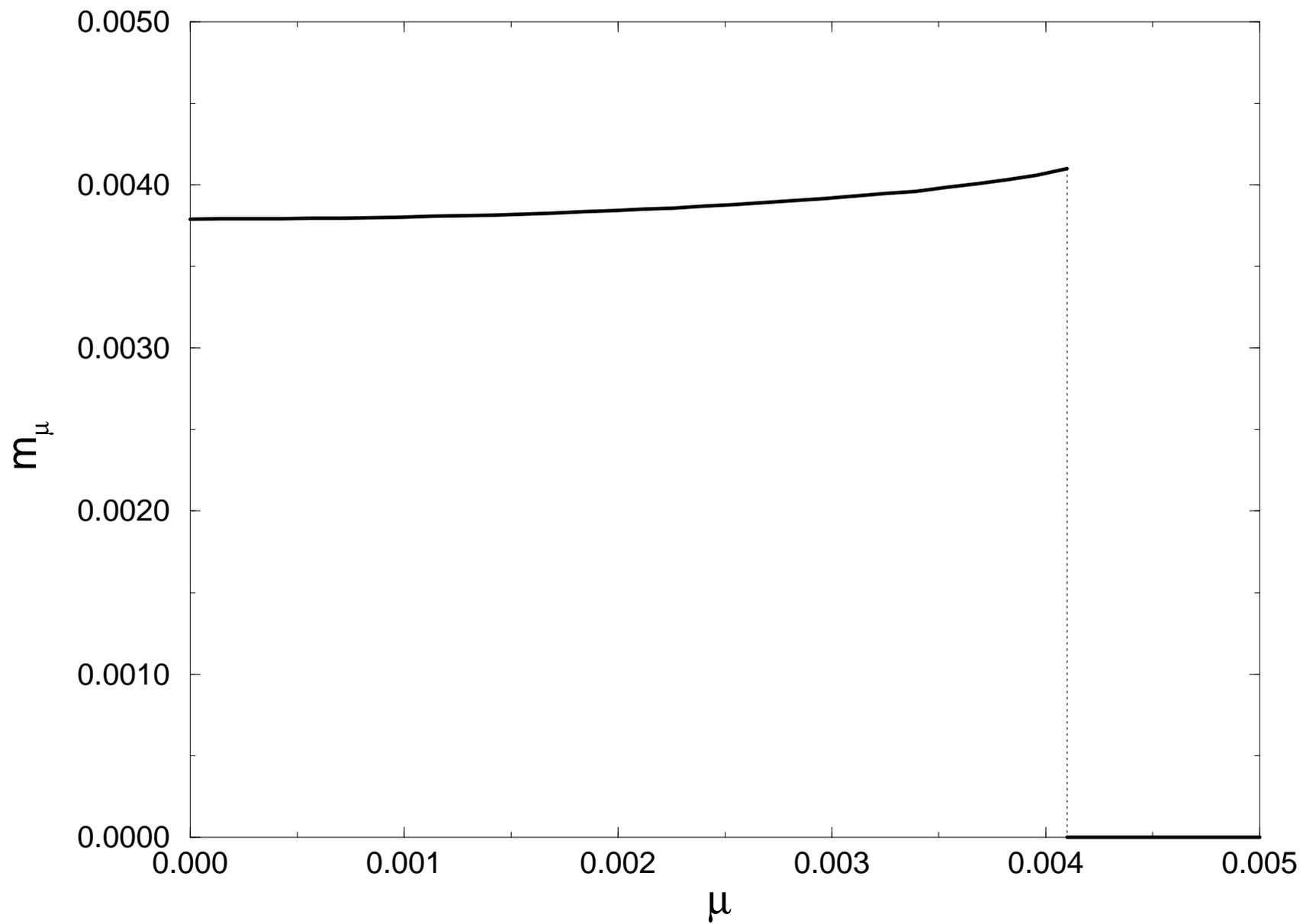
{\bf Fig. 2:}  The dynamical fermion mass $~m_{\mu}~$ as a function 
of the chemical potential $~\mu~$ (both measured in units of 
$\sqrt{2 |eH|}$) for $~\alpha = \pi/30$.  The corresponding 
dimensionless critical chemical potential $~\hat{\mu}_c~$ is 
equal to $~0.004101$.\\

\noindent
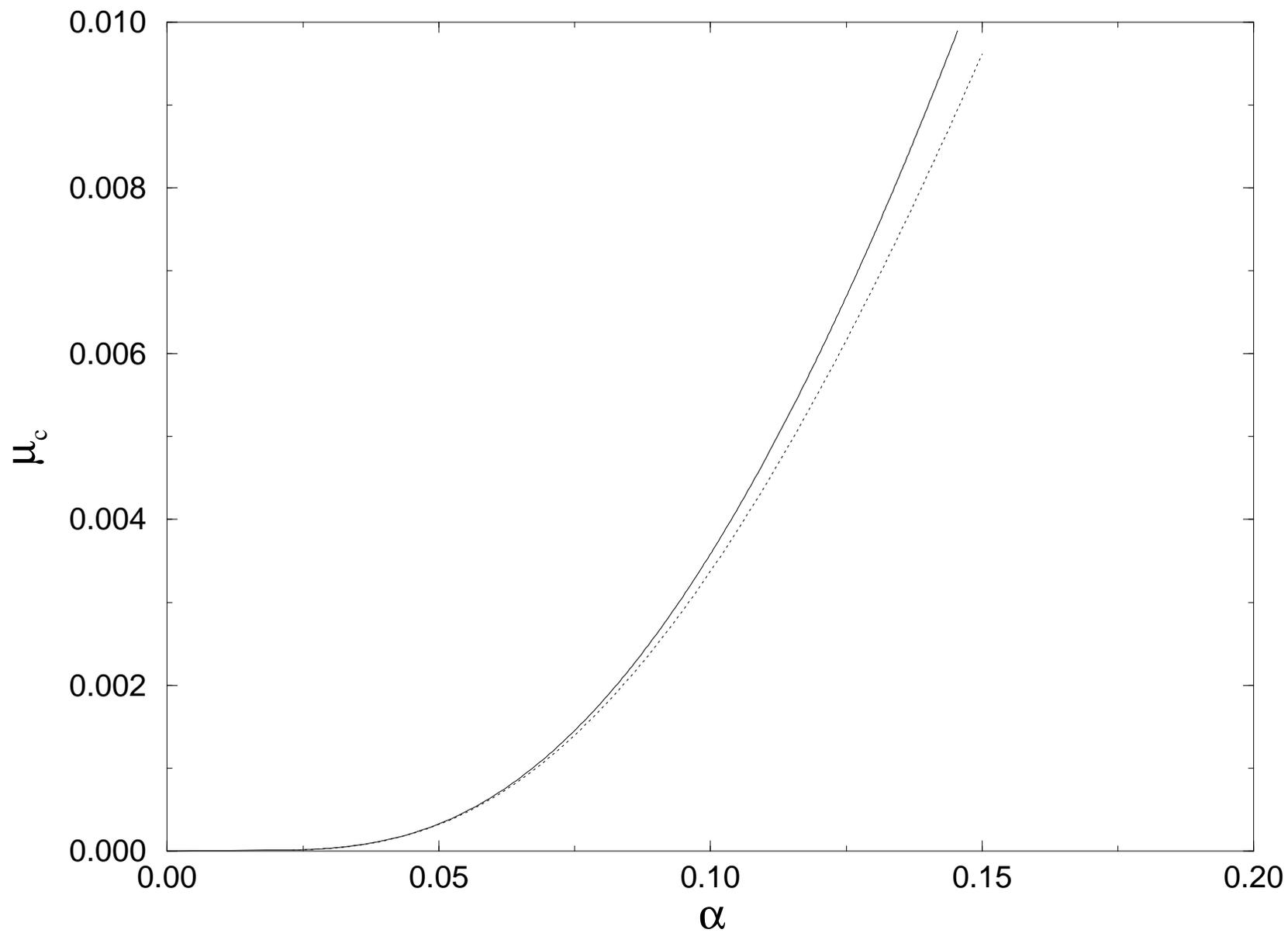
{\bf Fig. 3}:  The critical chemical potential (measured in units 
of $~\sqrt{2 |eH|}~$) as a function of the gauge coupling $~\alpha$.  
Solid curve: from exact numerical calculations.  Dotted curve: 
from the estimate in Eq.(\ref{mucrit}).\\

\noindent
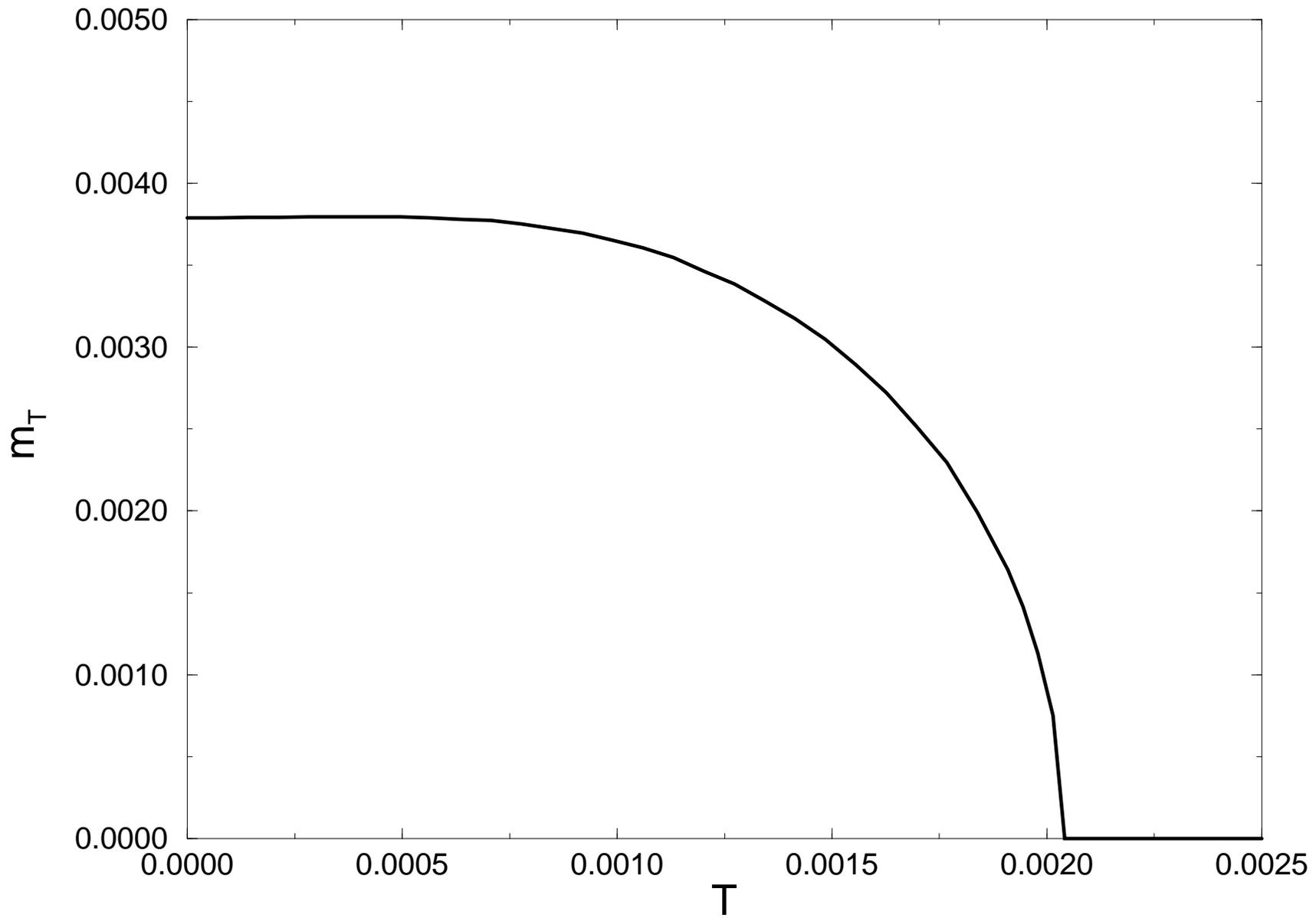
{\bf Fig. 4:}  The dynamical fermion thermal mass $~m_{T}~$ as a 
function of the temperature (both measured in units of 
$\sqrt{2 |eH|}$) for $~\alpha = \pi/30$.


\newpage
\begin{references}
\bibitem{Ng}
Y.J. Ng and Y. Kikuchi, in {\it Vacuum Structure in Intense 
Fields}, eds. H. M. Fried and B. M\"{u}ller (Plenum, New York, 
1991); Y.J. Ng, in {\it Tests of Fundamental Laws in Physics}, 
eds. O. Fackler and J. Tran Thanh Van (Editions Frontiers, 
33 Gif-sur-Yvette Cedex, 1989). 

\bibitem{GMS}
V. P. Gusynin, V. A. Miransky, and I. A. Shovkovy, Phys. Rev. 
D {\bf 52}, 4747 (1995); Nucl. Phys. {\bf B462}, 249 (1996).  
See also D. K. Hong, Y. Kim and S.-J. Sin, Phys. Rev. D 
{\bf 54}, 7879 (1996). 

\bibitem{LNA}
C. N. Leung, Y. J. Ng, and A. W. Ackley, Phys. Rev. D {\bf 54}, 
4181 (1996).  

\bibitem{LLN}
D.-S. Lee, C. N. Leung and Y. J. Ng, Phys. Rev. D {\bf 55}, 6504
(1997). 

\bibitem{GS}
V. P. Gusynin and I. A. Shovkovy, Phys. Rev. D {\bf 56}, 5251 
(1997).

\bibitem{Ka}
See, e.g., J. Kapusta, Finite-Temperature Field Theory (Cambridge 
University Press, Cambridge, England, 1989).

\bibitem{OC}
See Ref. 9 in Ref. \cite{LLN}.                  

\bibitem{typo}
See Eq.(64); there is a typographical mistake: 
${\rm e}^{-\sqrt{\frac{Q_{1,2}}{T}}}$ should be replaced by 
${\rm e}^{-\frac{\sqrt{Q_{1,2}}}{T}}$.

\bibitem{EXP}
In fact, one has to compute the corresponding effective potential 
to see if these solutions are stable and if the other branch of 
solutions corresponding to smaller values of $\hat{m}_{\mu}$ are 
unstable.  Here we content ourselves by simply applying the 
continuity criterion. 

\end{references}
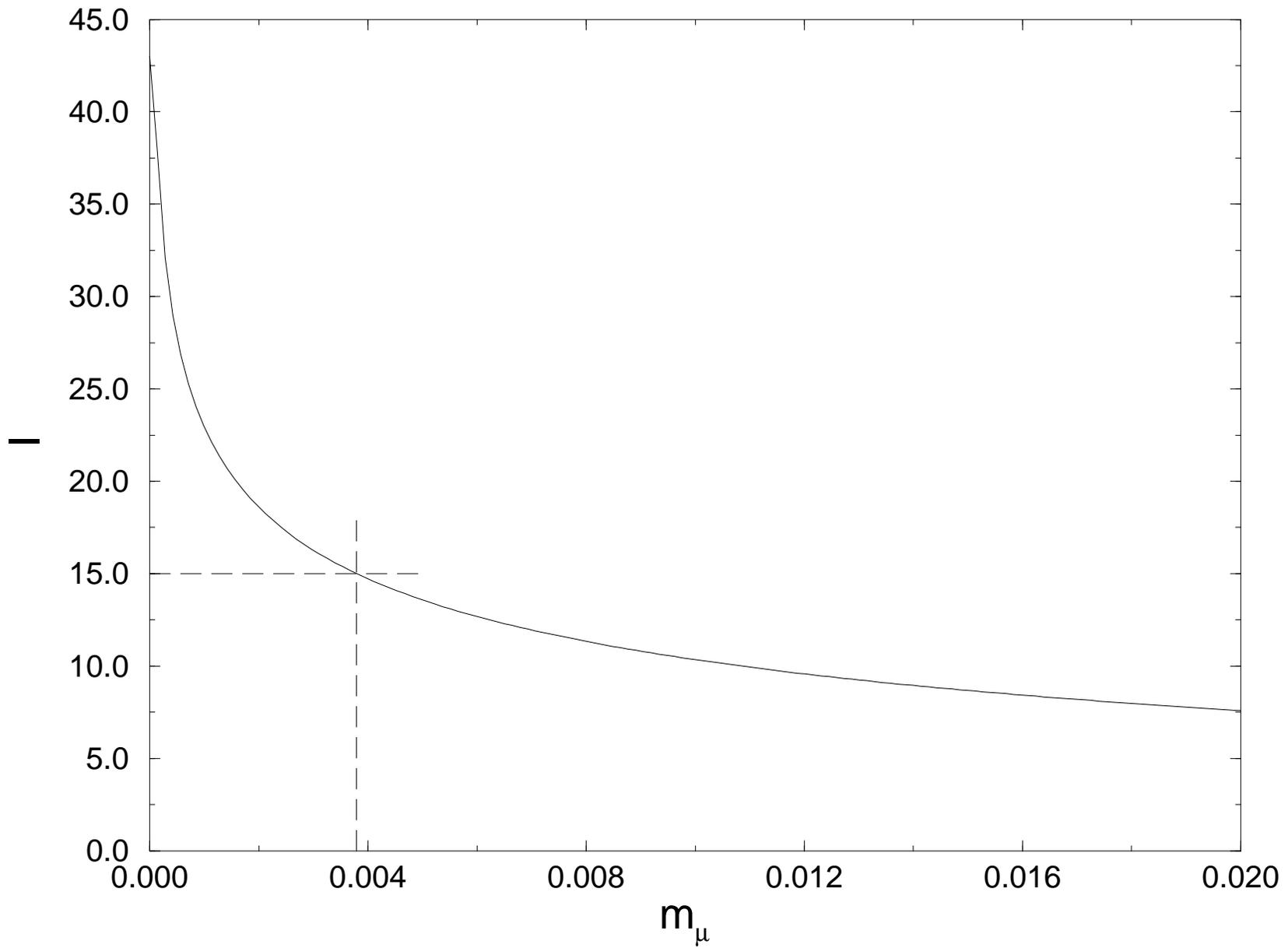
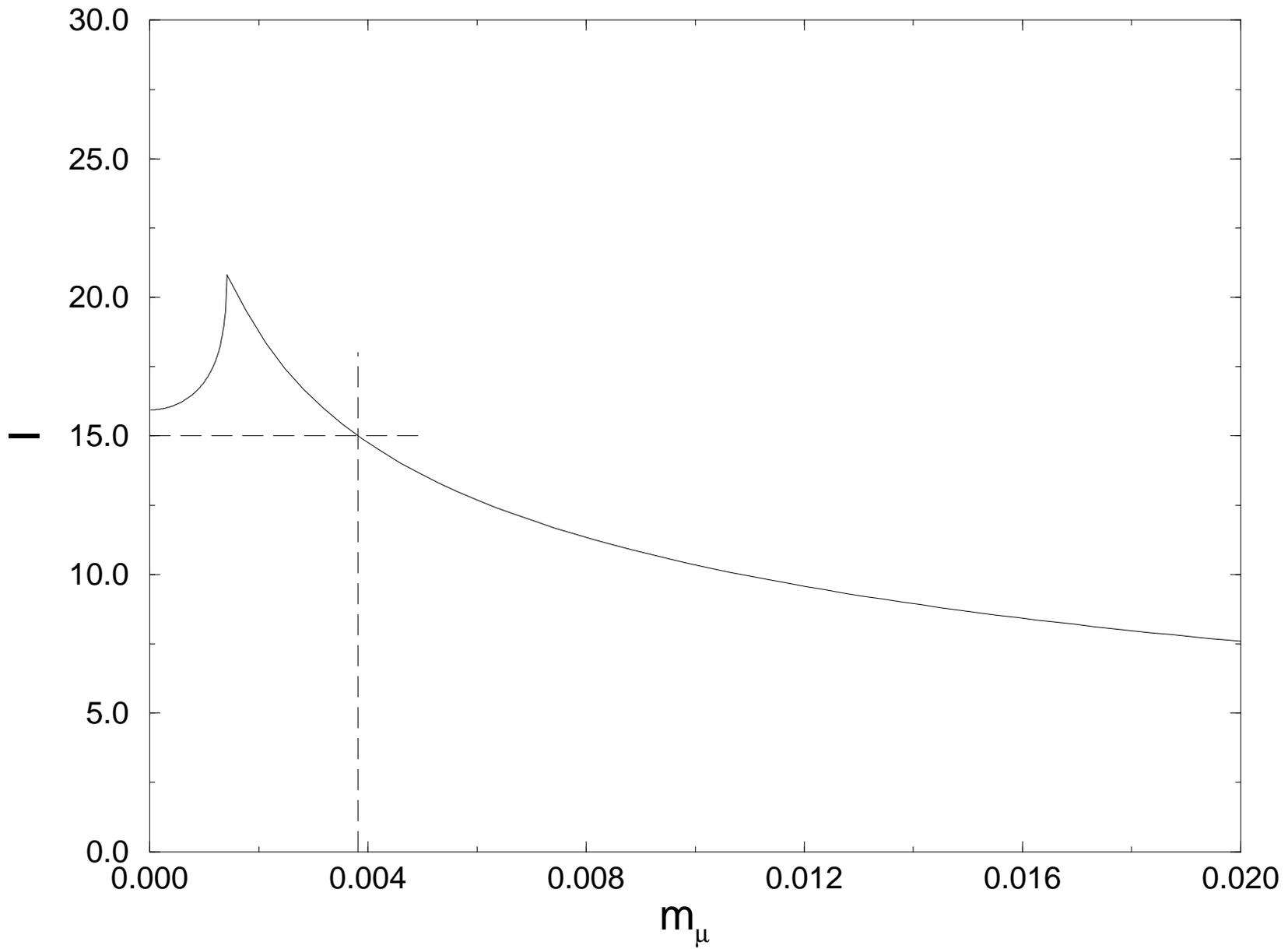
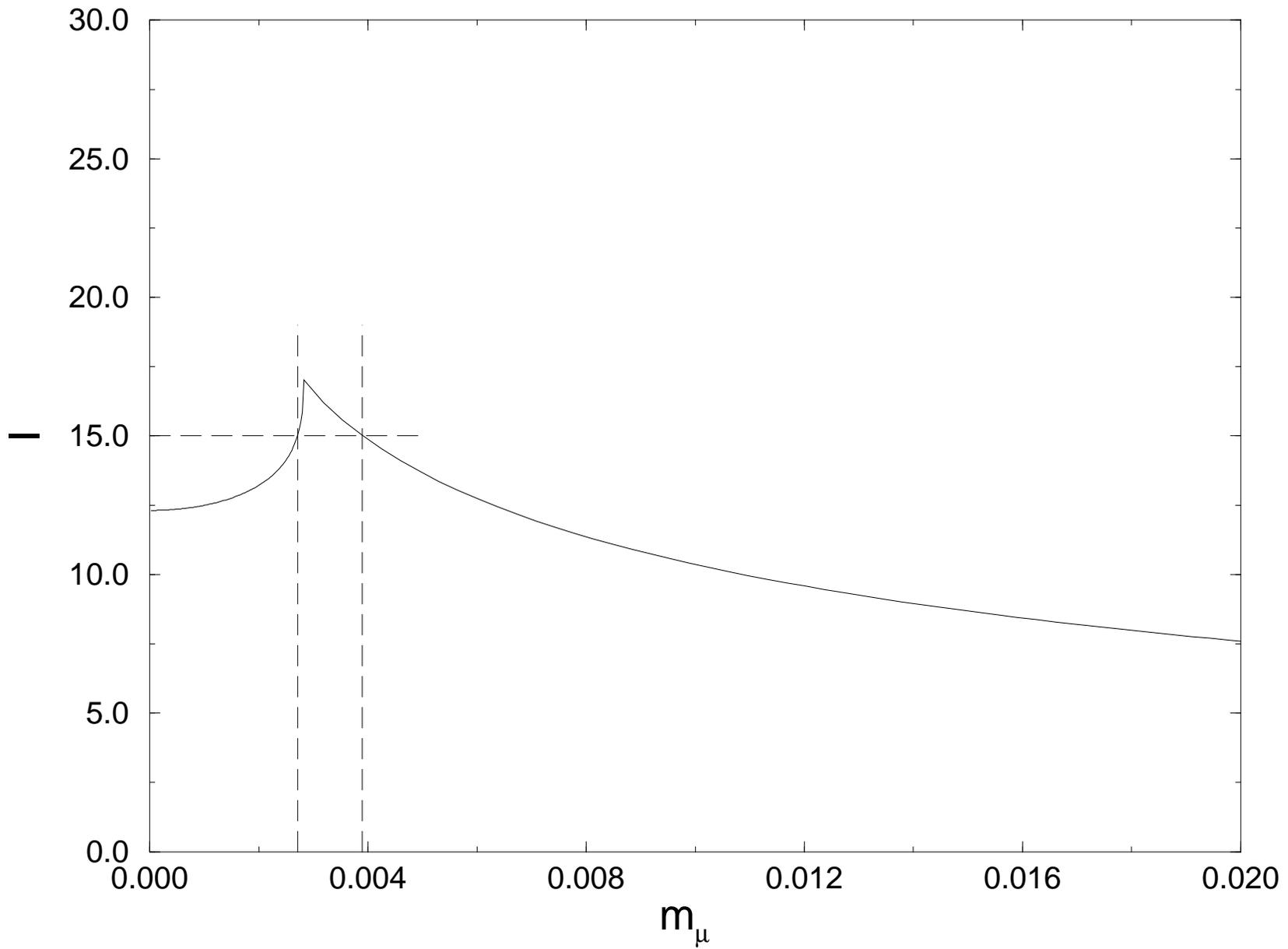
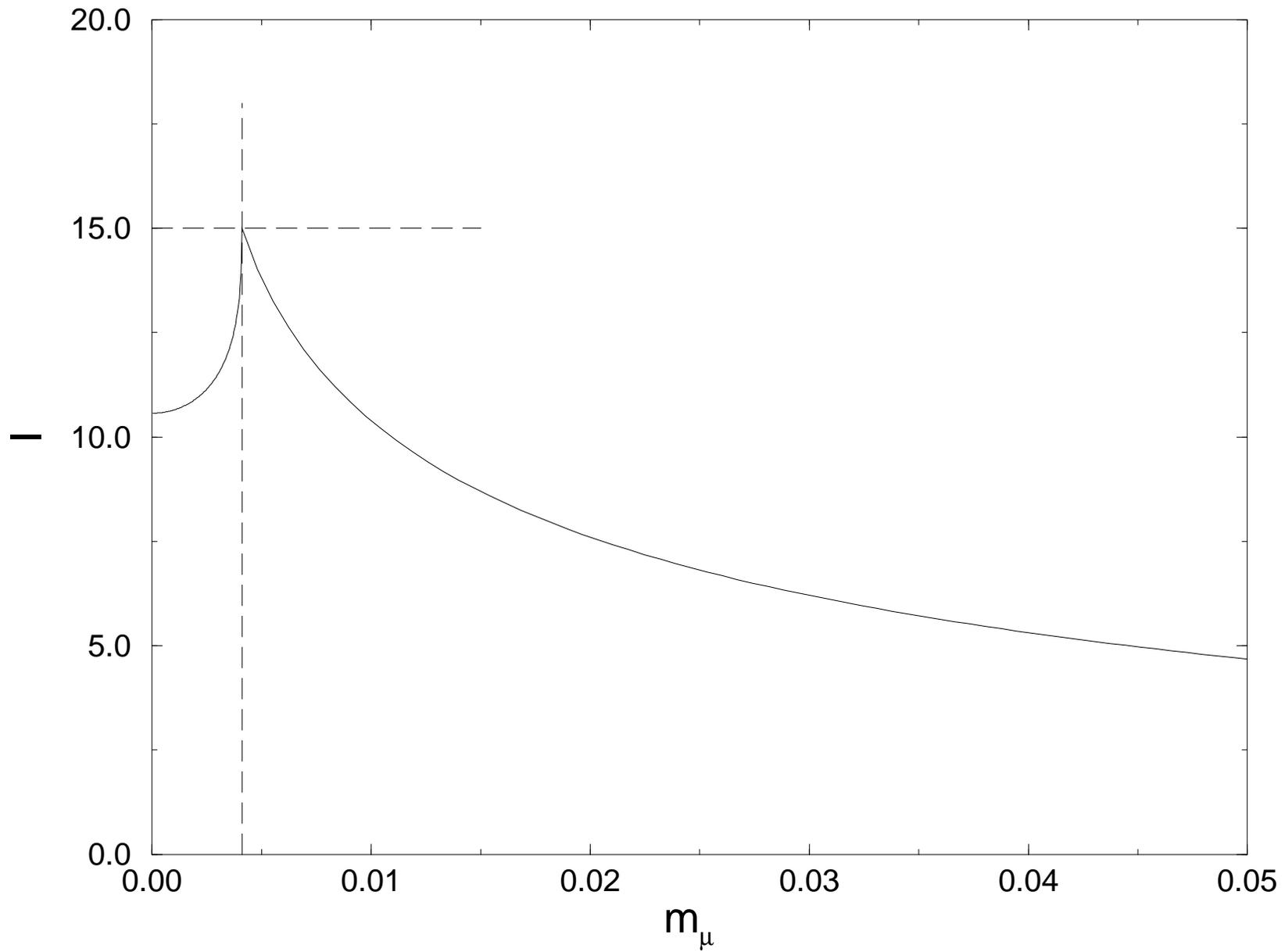
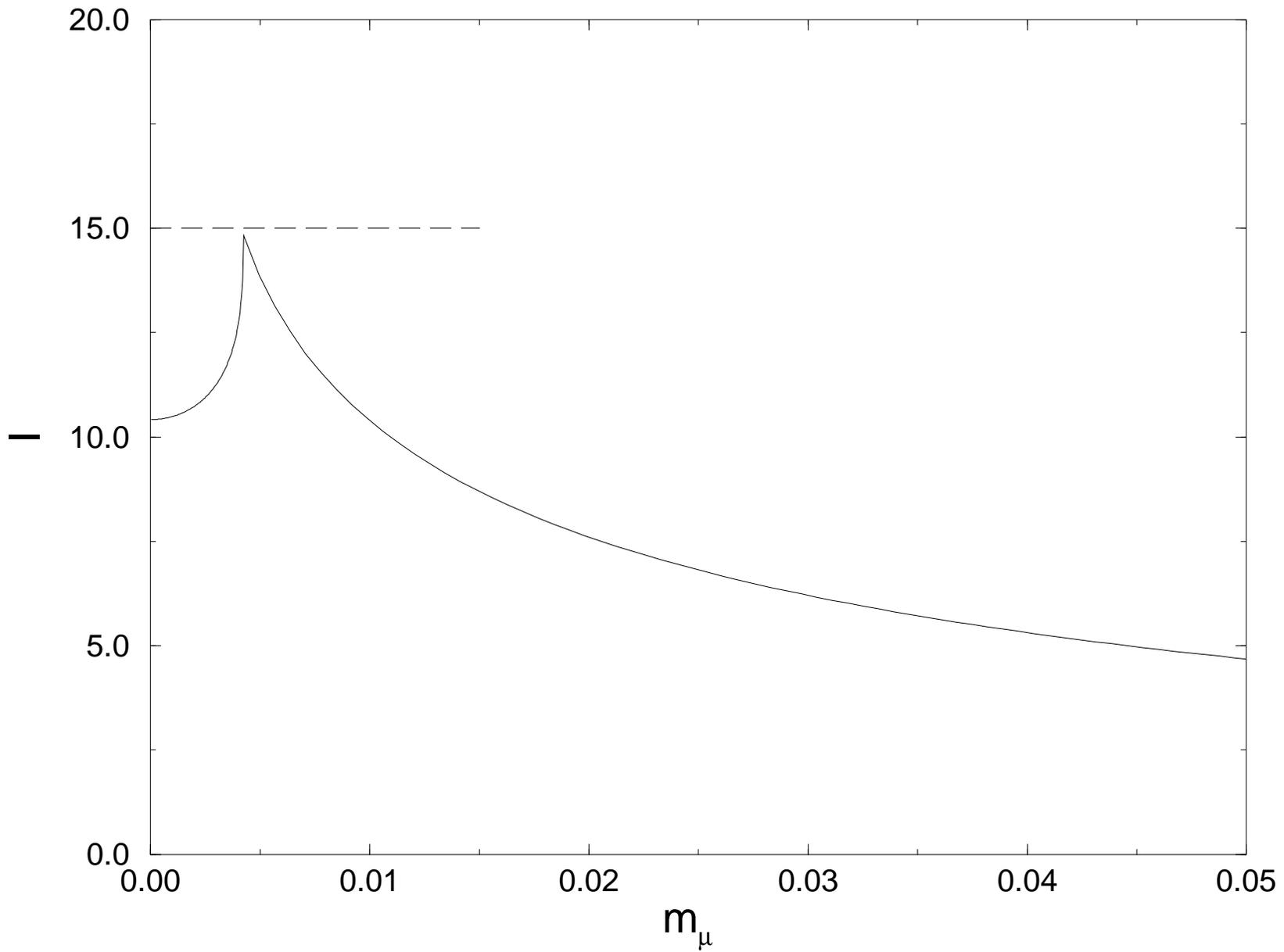
\end{document}